\documentclass[aps,prb,twocolumn,groupedaddress,showpacs]{revtex4}

\usepackage[]{graphicx}
\usepackage[]{psfig}

\begin{document}

%\preprint{}

\title{Monte Carlo Method for a Quantum Measurement Process by a 
Single-Electron Transistor}


\author{Hsi-Sheng Goan}
\email[E-mail: ]{goan@physics.uq.edu.au}
\thanks{Mailing address: Center for Quantum Computer
Technology, C/- Department of Physics, University of Queensland,
Brisbane 4072 Australia}
\affiliation{Center for Quantum Computer Technology, 
University of New South Wales,
Sydney, NSW 2052 Australia}


\date{\today}% It is always \today, today,

\begin{abstract}
We derive the quantum trajectory or stochastic (conditional) master
equation for a single superconducting Cooper-pair box (SCB) 
charge qubit measured by a single-electron transistor (SET) detector.
This stochastic master equation describes the random evolution of the
measured SCB qubit density matrix which both conditions and is conditioned on
a particular realization of the measured electron tunneling events
through the SET junctions.
Hence it can be regarded as 
a Monte Carlo method that allows us to simulate the
continuous quantum measurement process.  
We show that the master equation for the 
``partially'' reduced density matrix 
[Y.~Makhlin et.al., Phys. Rev. Lett. {\bf 85}, 4578 (2000)]
can be obtained
when a ``partial'' average is taken on the stochastic 
master equation over the fine grained measurement records of
the tunneling events in the SET. 
Finally, we present some Monte Carlo simulation results
for the SCB/SET measurement process. We also analyze 
the probability distribution $P(m,t)$ of
finding $m$ electrons that have 
tunneled into the drain of the SET in time $t$ to demonstrate the connection
between the quantum trajectory approach and the ``partially'' reduced
density matrix approach.
\end{abstract}

%\pacs{85.35.Be,03.67.Lx,73.63.Kv,73.63.Rt,73.23Hk,03.65.Bz,05.40Ca}
% PACS, the Physics and Astronomy Classification Scheme.
\maketitle

%\section{Introduction \label{sec:intro}}

\section{Introduction \label{sec:Intro}}

The single-electron transistor (SET) is a highly charge-sensitive
electro-meter and has been suggested as a readout device for
solid-state charge qubits \cite{Shnirman98,Makhlin00} 
or spin qubits \cite{Loss98,Kane98} (through a measurement of a
spin-dependent charge transfer).  
The problem of a charge qubit subject to a measurement by a SET
has been extensively studied in Refs.\ 
\onlinecite{Shnirman98} and \onlinecite{Makhlin00}.
We refer to the approach of these papers as
the master equation method of the ``partially'' reduced density matrix.
In this approach, one takes a trace over
environmental (detector) microscopic degrees of the freedom but 
keeps track of the number of electrons, $m(t)$, that have tunneled
through the SET into the drain during time $t$ in the 
``partially'' reduced density matrix.
If experimentally the number of accumulated electrons or current 
passing through the SET is measured, 
this approach can provide us with information about 
the initial qubit state. 
But, the system dynamics in this approach is still deterministic; 
i.e., this approach is still in an ensemble and time average sense.
%Hence, it cannot describe
%the conditional dynamics of the qubit system in a single realization of
%continuous measurements, which reflects the stochastic nature of
%electrons tunneling through the SET junctions.

A Monte Carlo method \cite{Bakhvalov89} which allows one to follow
each electron tunneling event has been successfully applied to
simulate transport properties of a SET or more
complicated single electronics circuits. 
This method gives physical insight into the processes taking place in the 
simulated system. But to our knowledge, it has not yet been formally applied to
quantum measurement problems by a SET detector.
In this paper, we provide such an investigation. 
We derive the {\em quantum-jump}  
stochastic master equation (or quantum trajectory equation) for a single
superconducting Cooper-pair box (SCB) charge qubit 
(generalization to other charge qubit case is simple)
continuously measured by a SET. 
This stochastic master equation describes the random evolution of the
measured SCB qubit density matrix which both conditions and is conditioned on
a particular realization of the measured electron tunneling events
through the SET junctions.
We can regard it as 
a Monte Carlo method that allows us to simulate the
continuous quantum measurement process of a charge qubit by a SET.  
This quantum trajectory approach (or Bayesian formalism) was introduced
recently \cite{Korotkov99,Goan01,Goan03} to describe a charge qubit
measured by a low-transparency point contact detector.
Here we present 
%to our knowledge for the first time, 
the {\em quantum-jump} stochastic master equation for the SET detector.
Especially, we show that the master equation for the 
``partially'' reduced density matrix (a ``partial'' course-grain description)
presented in Refs.~\onlinecite{Shnirman98}
and \onlinecite{Makhlin00} can be obtained by 
taking a ``partial'' average on the stochastic 
master equation over the fine grained measurement records of the
tunneling events in the SET. 
Finally, we present some
Monte Carlo simulation results for the SCB/SET measurement process.
We also analyze an important ensemble quantity for an initial
qubit state readout experiment, 
$P(m,t)$ the probability distribution \cite{Shnirman98,Makhlin00} of
finding $m$ electrons that have 
tunneled into the drain of the SET in time $t$. This analysis 
demonstrates further the connection
between the quantum trajectory approach presented here
and the ``partially'' reduced
density matrix approach in Refs.~\onlinecite{Shnirman98}
and \onlinecite{Makhlin00}.

\section{Model Hamiltonian \label{sec:Ham}}
The Hamiltonian of the SCB/SET system is described in Refs.\ 
\onlinecite{Shnirman98}
and \onlinecite{Makhlin00} as:
\begin{equation}
{\cal H}={\cal H}_{\rm SET} +{\cal H}_{\rm L}+{\cal H}_{\rm R}+{\cal
  H}_{\rm I} +{\cal H}_{\rm T} +{\cal H}_{\rm qb}+{\cal H}_{\rm int}.
\label{H}
\end{equation}
Briefly, 
\begin{equation}
{\cal H}_{\rm SET}=E_{\rm SET}(N-N_g)^2
\end{equation}
describes the charging energy of the SET.
The charge on the middle island is $e N$, and 
the induced charge $e N_g$ is determined by the gate voltage $V_{\rm g}$ 
and other voltages in the circuit.
\begin{equation}
  \label{eq:LRI}
{\cal H}_{\rm r}=\sum_{ks}\epsilon_{ks}^r
c_{ks}^{r\dagger}c_{ks}^{r},  
\end{equation}
where $r={L,R,I}$,
describes microscopic degrees of freedom of noninteracting electrons
in the two leads (left and right) and the middle island of the SET,
respectively.   
To make the charge transfer explicit, 
two ``macroscopic''  operators, $e^{\pm i\phi}$ and $e^{\pm i\psi}$
are included in the tunneling Hamiltonian  \cite{Shnirman98,Makhlin00} 
in the SET:
\begin{eqnarray}
{\cal H}_{\rm T} = 
&& 
\sum_{kk's} T^{L}_{kk's}
        c^{L\dagger}_{ks} c_{k's}^{I}
        e^{-i\phi}
\nonumber \\ 
&+&
   \sum_{k'k''s}  T^{R}_{k'k''s} c^{
        R\dagger}_{k''s}  c_{k's}^{I}
        e^{-i\phi} e^{i\psi} + {\rm H.c.} \ .
\label{HT}
\end{eqnarray} 
The effective Hamiltonian of the uncoupled qubit, written in the
charge eigen basis of the number $n$ of extra Cooper pair on the island
of SCB, is
\begin{equation}
  \label{eq:qb}
H_{\rm qb}=\frac{1}{2}(E_{\rm ch}\hat\sigma_z-E_J\hat\sigma_x),  
\end{equation}
where $\hat n=(1-\hat\sigma_z)/2$ with eigenvalues $n=0$ or $1$.
The capacitive Coulomb coupling between the charge on the SET island
and that on the SCB qubit is represented by
\begin{equation}
  \label{eq:Hint}
  {\cal H}_{\rm int}=2E_{\rm int}N \hat n \, .
\end{equation}

We will consider the case that the leading
tunneling process in the SET are sequential transitions between two
adjacent charge states $N$ and $N+1$ (say, $N=0$ and  $N+1=1$ states
to represent the extra charge on the SET island).  
This would be the case if the applied transport voltage across the SET
is not too high and 
the temperature is low (for simplicity, we consider the zero
temperature case).
Since, effectively, only two
adjacent charge states $N=0$ and  $N+1=1$ are considered,  
the charge transfer operators $e^{\pm i \phi}$, in this case, 
satisfy $e^{-i\phi}|N\rangle=0= e^{i\phi}|N+1\rangle$,
$e^{i\phi}|N\rangle=|N+1\rangle$ and $e^{-i\phi}|N+1\rangle=|N\rangle$.
The other set of charge transfer operators satisfy
$e^{\pm i\psi}|m\rangle=|m\pm 1 \rangle$,
where $m$ represents 
the number of electrons that has tunneled into the right lead (drain)
of the SET.

\section{Measurement records and conditional density matrix \label{sec:ME}}

To be able to describe the measured qubit in a pure state continuously, 
one needs to have the maximum knowledge about the change of its state.
When the qubit interacts with (is measured by) the SET, 
this information is lost to the SET. 
For example, each time when an electron tunnels onto or off the
SET island, it will cause a change (e.g., a phase shift) 
of the qubit state. One
can recover this information lost, provided that 
a detailed measurement record from the SET is available. 
The transport of electrons through the SET occurs via real states of
the central island, from $N\to N+1\to N$. 
The information of detecting the $m^{\rm th}$ electron just
tunneling into the drain only tells us that the island state now is
in $|N\rangle$ state. 
Thus knowing the ``partially'' reduced density matrix $\rho_{N}(m,t)$
at every time \cite{Shnirman98,Makhlin00}  
does not provide us with the full information.

One can imagine that 
in the transport process, electrons may spend different times 
in the intermediate $|N+1\rangle$ state, 
causing different phase shifts to the qubit. 
If the record of
the times when electrons tunneling onto and off the island is not 
available from the measurement results of the SET, our knowledge about
the precise qubit state decreases. 
When this happens, averaging 
the random dwelling times of electrons on the island 
over a period of time or over an ensemble of systems
will then lead to the decoherence of the qubit. 
Hence, one needs to have a measurement record which records  
whether or not an electron tunnels onto or off the central island of the SET at
each time interval $dt$. This time interval $dt$ should be 
much smaller than the typical qubit system evolution or response time
so that no information is lost as far as the qubit system evolution is
concerned. In this sense, effectively the qubit is continuously
monitored or measured.

For this purpose, we introduce $dN_{Lc}(t)$ and $dN_{Rc}(t)$ to
represent, in the {\em quantum-jump} case \cite{Goan01}, the number 
(either zero or one) of 
tunneling events seen in infinitesimal time $dt$ 
through the left and right junctions of the SET, respectively. 
Throughout the paper, the subscript or superscript $c$ indicates that
the quantity to which it is attached is conditioned on previous
measurement results.
If no tunneling electron is detected, the result is {\em null}, 
i.e., $dN_{Lc}(t)=0$ and $dN_{Rc}(t)=0$.
If there is {\em detection} of a tunneling electron in time $dt$, 
then $dN_{Lc}(t)=1$ or $dN_{Rc}(t)=1$.
We can think of $dN_{Rc}(t)$
as the increment in the number of electrons
$N_{Rc}(t)=\sum dN_{Rc}(t)$ passing through the right junction of the
SET into the drain
in the infinitesimal time $dt$. 
It is the variable $N_{R}(t)=m(t)$, the accumulated
electron number transmitted through the SET in
the drain, which is used in Refs.\
\onlinecite{Shnirman98}
and \onlinecite{Makhlin00}.  
Since the nature of detection results is classical and that of  
electrons tunneling through the SET is stochastic, 
$dN_{Lc/Rc}(t)$ should represent a classical random process. 
The measurement record in each single run of experiment
is the set of times $\{t_L^{(i)}\}$ and $\{t_R^{(i)}\}$ when electrons
tunnel onto or off the SET island, respectively 
[i.e., ones of $dN_{Lc}(t)$ and $dN_{Rc}(t)$ over the entire detection
time; see, e.g., Fig.~\ref{fig:traj}(i) and (j)].

At first, one may expect that at the end of each time interval $dt$,
there are four possible measurement outcomes, $dN_{Lc}(t) \, [1-dN_{Rc}(t)]$, 
$dN_{Rc}(t) \, [1-dN_{Lc}(t)]$, $[1-dN_{Lc}(t)] \, [1-dN_{Rc}(t)]$, 
and $dN_{Lc}(t)\, dN_{Rc}(t)$. It is important to realize that a null
result (e.g., $dN_{Lc}=dN_{Rc}=0$) in a time interval $dt$ is still a
measurement result or outcome.  
Let us consider the case in the sequential tunneling
dominated regime   
that the probability of electrons tunneling onto and off the SET 
island within the same infinitesimal time interval $dt$ is rather
small. In fact, the respective
probability of $dN_{Lc}(t)$ or $dN_{Rc}(t)$ equal to unity is
proportional to $dt$ [see Eqs.~(\ref{dNL}) and (\ref{dNR})]. 
Thus the product of $dN_{Lc}(t)\, dN_{Rc}(t)=1$ occurs with
probability proportional to $dt^2$.
Since we shall keep only terms to order $dt$ in the master equations,
we can neglect the case that $dN_{Lc}(t)$ and $dN_{Rc}(t)$ both equal 
one within the same infinitesimal time interval. 
The possible measurement outcomes then become: $dN_{Lc}(t)$,
$dN_{Rc}(t)$ and $[1-dN_{Lc}(t)-dN_{Rc}(t)]$.
The first two terms, in this case, 
represent that an electron tunneling
event through, respectively, the left and right junctions of SET is measured 
at the end of the time interval [$t,t+dt$).
While the last term $[1-dN_{Lc}(t)-dN_{Rc}(t)]$ represents 
that no tunneling event is observed in [$t,t+dt$). 
Thus the conditioned density matrix $W_c(t+dt)$ to order $dt$ 
at the end of the time interval [$t,t+dt$)
can be written as 
\begin{eqnarray}
W_c(t+dt)&=&dN_{Lc}(t) \frac{W_{L1c}(t+dt)}{{\rm Tr}[W_{L1c}(t+dt)]}
\nonumber\\
&&\hspace*{-1cm}
+dN_{Rc}(t) \frac{W_{R1c}(t+dt)}{{\rm Tr}[W_{R1c}(t+dt)]}
\nonumber\\
&&\hspace*{-1cm}
+[1-dN_{Lc}(t)-dN_{Rc}(t)] \frac{W_{0c}(t+dt)}{{\rm Tr}[W_{0c}(t+dt)]},
\label{condDM}
\end{eqnarray}
where $W_{L1c}(t+dt)$, $W_{R1c}(t+dt)$, and $W_{0c}(t+dt)$ 
are the unnormalized density matrices, given that an electron tunneling
event through left or right junction of the SET island, or no
tunneling event is measured at the end of the time interval [$t,t+dt$). 
Equation (\ref{condDM}) simply states that when $dN_{Lc}=1$ and
$dN_{Rc}=0$, the normalized 
conditioned density matrix is $W_{L1c}(t+dt)/{\rm Tr}[W_{L1c}(t+dt)]$,
and so on.  
Self-consistently, the ensemble
averages $E[dN_{Lc}(t)]$ and $E[dN_{Rc}(t)]$ of the classical
stochastic processes $dN_{Lc}(t)$ and 
$dN_{Rc}(t)$ should 
equal respectively
the probabilities (quantum average) of 
electrons tunneling through the left and right junctions of
the SET island in time $dt$, i.e., 
$E[dN_{Lc}(t)]={\rm Tr}[W_{L1c}(t+dt)]$ and 
$E[dN_{Rc}(t)]={\rm Tr}[W_{R1c}(t+dt)]$.

Formally, we can write the currents through the junctions as 
\begin{eqnarray}
  \label{eq:currentL}
I_{Lc}(t)=e\, [{dN_{Lc}(t)}/{dt}],
\\
I_{Rc}(t)=e\, [{dN_{Rc}(t)}/{dt}]. 
\label{eq:currentR} 
\end{eqnarray}
The question now is to find expressions for 
$W_{L1c}(t+dt)$, $W_{R1c}(t+dt)$, and $W_{0c}(t+dt)$ in the model.
To do this, we 
derive the unconditional master equation and then use it to find 
$W_{L1c}(t+dt)$, $W_{R1c}(t+dt)$, and $W_{0c}(t+dt)$.

\section{Stochastic master equation \label{sec:SME}}

Following the same assumptions and approximations in 
Refs.\ \onlinecite{Shnirman98}
and \onlinecite{Makhlin00} and similar derivations
in Refs.\ \onlinecite{Goan01} and \onlinecite{Wiseman01},
we first derive the master equation of ``partially'' reduced density matrix.
By tracing out the microscopic degrees of freedom of the left and
right leads and the island of the SET, 
but keeping the electron transfer 
operators explicitly  \cite{Shnirman98,Makhlin00} 
(so that we can keep track of effects of 
electron tunneling events on the system density matrix), 
we obtain the Born-Markov master equation for
the ``partially'' reduced density matrix operator ${\cal W}(t)$
of the SCB/SET system (consisting of the qubit, and the island and drain
of the SET) as:
\begin{eqnarray} 
[{d{\cal W}(t)}/{dt}]&=&-({i}/{\hbar})
[{\cal H}_{\rm qb}+{\cal H}_{\rm int}, {\cal W}(t)]
\nonumber \\
&&\hspace*{-1.3cm}+
\Gamma_{L} {\cal D}[e^{i\phi}(1-\hat{n})]{\cal W}(t) 
+\Gamma'_{L} {\cal D}[e^{i\phi}\hat{n}]{\cal W}(t)
\nonumber \\
&&\hspace*{-1.3cm}
+\Gamma_{R} {\cal D}[e^{-i\phi}e^{i\psi}(1-\hat{n})]{\cal W}(t) 
+\Gamma'_{R} {\cal D}[e^{-i\phi}e^{i\psi}\hat{n}]{\cal W}(t)
\nonumber \\
&&\hspace*{-1.3cm}-{(\Gamma_{L}+\Gamma'_{L})} 
[\hat{n},[\hat{n},e^{i\phi}\,{\cal W}(t)\,e^{-i\phi}]]/2
\nonumber \\
&&\hspace*{-1.3cm}-{(\Gamma_{R}+\Gamma'_{R})} 
[\hat{n},[\hat{n},e^{-i\phi}e^{i\psi}\,{\cal W}(t)\,e^{-i\psi}e^{i\phi}]]/2
\; ,
\label{MEpartial}
\end{eqnarray} 
where ${\cal D}$ is defined 
for arbitrary operators $B$ and ${\cal W}$ as
\begin{eqnarray}
  \label{eq:D}
{\cal D}[B]{\cal W}&=&{\cal J}[B]{\cal W}-{\cal A}[B]{\cal W},
\\
{\cal J}[B]{\cal W} &= &B {\cal W} B^\dagger,
\\
{\cal A}[B]{\cal W} &=& (B^\dagger B {\cal W} +{\cal W} B^\dagger B)/2. 
\end{eqnarray}
The rates $\Gamma_{L/R}$ and $\Gamma'_{L/R}$ represent
the tunneling rates (in the left or right junction)
with and without the presence of the extra Cooper pair 
on the island of the SCB (i.e., $n=1$ or $n=0$), respectively.
They are determined by the chemical potentials $\mu_{L/R}$ of
the leads and the induced charge $N_g$ on the SET's island:
\begin{eqnarray}
\Gamma_{L} &=& (2\pi\alpha_{L}/\hbar)[\mu_{L}-(1-2N_g) E_{\rm SET}],
\label{GammaL}\\
\Gamma_{R} &=& (2\pi\alpha_{R}/\hbar)[(1-2N_g)E_{\rm SET}-\mu_{R}],
\label{GammaR}\\
\Gamma'_{L} &=& \Gamma_{L}-(4\pi\alpha_{L}E_{\rm int}/\hbar),
\label{GammaL'}\\
%and 
\Gamma'_{R} &=& \Gamma_{\rm R}+(4\pi\alpha_{R}E_{\rm int}/\hbar),
\label{GammaR'}
\end{eqnarray}
where $\alpha_{L/R}=R_Q/(4\pi^2 R_{L/R})$, $R_Q=h/e^2$ is the
resistance quantum, 
%$R^{-1}_{L/R}=(4\pi^2 e^2/h) g_I g_{L/R}|T^{L/R}|^2$ 
and $R_{L/R}$ represents
the resistance of the left or right junction.

The master equation (\ref{MEpartial})
has a translational symmetry in $m$ space. So
by summing all possible values of
$m$ of the right reservoir (drain) states completely, 
a closed form of the master equation of  
$W(t)=\sum_m\langle m|{\cal W}(t)|m\rangle$ can be obtained. 
The resultant equation is equivalent to Eq.\ (\ref{MEpartial}) but with
the replacements of $e^{\pm i\psi}\to 1$ and 
${\cal W}(t)\to W(t)$.
One may expect to apply the similar sum 
procedure to the island states.  
However, 
since effectively only two extra adjecent charge states
$|N\rangle$ and $|N+1\rangle$ are considered, 
a closed form of the master equation for the qubit density matrix
operator alone, 
$\rho(t)=\langle N|W(t)|N\rangle+\langle N+1|W(t)|N+1\rangle
\equiv \rho_N(t)+\rho_{N+1}(t)$,
can not be obtained without further approximations, 
where $\rho_{N/N+1}(t)$ is each a $2\times 2$
operator in the qubit basis. 
One approach \cite{Wiseman01} is to assume extremely asymmetric 
tunnel junctions for the SET, i.e., one of the tunneling rates
through the junctions is much larger than the other. In this case 
one can apply the adiabatic elimination procedure \cite{Wiseman01}
to eliminate the
degrees of freedom of the SET island to obtain the reduced density
matrix for the qubit alone. But this asymmetric assumption is
equivalent to treating the SET as effectively a single junction
device, similar to a point contact detector. 
Here, however, we take the joint density matrix of the qubit and extra
charge on the SET island as the system density matrix in 
Eq.\ (\ref{condDM}).
After evaluating $W_c(t)$ [or 
$\rho^c_N(t)$ and $\rho^c_{N+1}(t)$] from 
the conditional master equation [see Eqs.\ (\ref{condMEN}) and
(\ref{condMEN+1})],
we can then find the conditional qubit density matrix operator alone
by writing $\rho^c(t)={\rm Tr}_N[W_c(t)]=\rho^c_N(t)+\rho^c_{N+1}(t)$.

Using the definition of the superoperator ${\cal D}$ and the fact that
the charge transfer operators are explicitly kept in
each term in
Eq.\ (\ref{MEpartial}), one can then find \cite{Goan01} from 
there [or more precisely from the master equation for $W(t)$]
the unnormalized density matrices, given that an electron tunneling
event through left or right junction of the SET island 
%or no tunneling event 
takes place at the end of the time interval
[$t,t+dt$), as: 
\begin{eqnarray}
W_{L1c}(t+dt)&=& dt \left\{ 
\Gamma_L {\cal J}[e^{i\phi}(1-\hat{n})]W_c(t)
\right.
\nonumber
\\ &&  \hspace*{-0.8cm}+\Gamma'_L {\cal J}[e^{i\phi}\hat{n}]W_c(t)
\nonumber
\\ &&\hspace*{-0.8cm} \left.
 -(\Gamma_L+\Gamma'_L)[\hat n, [\hat{n},
e^{i\phi}W_c(t)e^{-i\phi}]]/2\right\},
\label{LjumpDM}
\\
W_{R1c}(t+dt)&=& dt \left\{ 
\Gamma_R {\cal J}[e^{-i\phi}(1-\hat n)]W_c(t)
\right. \nonumber
\\ && \hspace*{-0.8cm}+\Gamma'_R {\cal J}[e^{-i\phi}\hat n]W_c(t)
\nonumber
\\ && \hspace*{-0.8cm}\left.
 -(\Gamma_R+\Gamma'_R)[\hat n, [\hat n,
e^{-i\phi}W_c(t)e^{i\phi}]]/2\right\}.
\label{RjumpDM}
\end{eqnarray}
It is required that the unconditional (ensemble averaged) density matrix 
$E[W_c(t+dt)]=W(t+dt)=W_{L1}(t+dt)+W_{R1}(t+dt)+W_{0}(t+dt)$. 
Hence we find $W_{0c}(t+dt)$, from the master equation for $W(t)$, as
\begin{eqnarray}
W_{0c}(t+dt)&=&W_c(t)
-dt({i}/{\hbar})[{\cal H}_{\rm qb}+{\cal H}_{\rm int},W_c(t)]
\nonumber
\\
&&\hspace*{-1.2cm}-dt \left\{\Gamma_L {\cal A}[e^{i\phi}(1-\hat
n)]+\Gamma'_L {\cal A}[e^{i\phi}\hat
n]
\right. \nonumber
\\
&&\hspace*{-1.2cm} \left. 
+\Gamma_R {\cal A}[e^{-i\phi}(1-\hat
n)]+\Gamma'_R {\cal A}[e^{-i\phi}\hat
n]\right\}W_c(t).
\label{nojumpDM}
\end{eqnarray}
Substituting Eqs.\ (\ref{LjumpDM})-(\ref{nojumpDM}) into Eq.\
(\ref{condDM}) and  replacing 
${\rm Tr}[W_{0c}(t+dt)]=1-{\rm Tr}[W_{L1c}(t+dt)]-{\rm Tr}[W_{L1c}(t+dt)]$, 
then keeping only the terms to order $dt$ in the resultant
equation  \cite{Goan01}, and finally evaluating this equation in $|N\rangle$ 
and $|N+1\rangle$ states respectively, 
we obtain the conditional master equation:   
\begin{eqnarray}
d\rho^c_N(t+dt)&=&-[dN_{Lc}(t)+dN_{Rc}(t)]\rho^c_N(t)
\nonumber
\\
&&+dN_{Rc}(t)\,[{\check\Gamma_R} \rho^c_{N+1}(t)/{{\cal P}_{R1c}(t)}]
\nonumber
\\
&&-dt\left\{({i}/{\hbar})[{\cal H}_{\rm qb}, \rho^c_N(t)]
+
\check\Gamma_L\rho^c_N(t)\right.
\nonumber
\\
&&\left.
-[{\cal P}_{L1c}(t)+{\cal P}_{R1c}(t)]\rho^c_N(t) \right\} \; ,
\label{condMEN}
\\
d\rho^c_{N+1}(t+dt)&=&-[dN_{Lc}(t)+dN_{Rc}(t)]\rho^c_{N+1}(t)
\nonumber
\\
&&+dN_{Lc}(t)\, [{\check\Gamma_L} \rho^c_{N}(t)/{{\cal P}_{L1c}(t)}]
\nonumber
\\
&&
-dt\left\{({i}/{\hbar})[
{\cal H}_{\rm qb}+2E_{int} \hat{n}, \rho^c_{N+1}(t)]
\right. 
\nonumber
\\
&&+\check\Gamma_R\rho^c_{N+1}(t)
\nonumber
\\
&&\left. -[{\cal P}_{L1c}(t)+{\cal P}_{R1c}(t)]\rho^c_{N+1}(t) \right\}
\; ,
\label{condMEN+1}
\end{eqnarray}
where ${\cal P}_{Lc}(t)$ and ${\cal P}_{Rc}(t)$ appearing in 
Eqs.\ (\ref{condMEN}) and (\ref{condMEN+1}) are due to the
normalization requirement for the density matrix 
after each detection interval $dt$ as in Eq.\ (\ref{condDM}),
and are given by
\begin{eqnarray}
{\cal P}_{L1c}(t)&=&\Gamma_{\rm L}{\rm Tr}[\rho^c_N(t)] 
+(\Gamma'_{\rm L}-\Gamma_{\rm L}) {\rm Tr}[\hat{n}\, \rho^c_{N}(t)] 
\; ,
\label{PL1}
\\
{\cal P}_{R1c}(t)&=&\Gamma_{\rm R}{\rm Tr}[\rho^c_{N+1}(t)] 
+(\Gamma'_{\rm R}-\Gamma_{\rm R}) {\rm Tr}[\hat{n}\, \rho^c_{N+1}(t)].
\label{PR1}
\end{eqnarray}
The rates $\check\Gamma_{L}$ and $\check\Gamma_{R}$
are defined as
\begin{eqnarray}
\check\Gamma_{L}\rho^c_{N}&=&\Gamma_{L}\rho^c_{N}+
{(\Gamma_{L}-\Gamma'_{L})}
\{\hat{n},\rho^c_{N}\}/2
\; ,\\
\check\Gamma_{R}\rho^c_{N+1}&=&\Gamma_{\rm R}\rho^c_{N+1}+
{(\Gamma_{R}-\Gamma'_{R})}
\{\hat{n},\rho^c_{N+1}\}/2
\; .
\end{eqnarray}
Self-consistently, $E[dN_{Lc}(t)]$ and $E[dN_{Rc}(t)]$
should equal their respective quantum averages, and 
from Eqs.\ (\ref{LjumpDM}) and (\ref{RjumpDM}) can be written 
 as \cite{phaseshift}
\begin{eqnarray}
E[dN_{Lc}(t)]&=&{\rm Tr} [W_{L1c}(t+dt)]={\cal P}_{L1c}(t)dt
\; ,
\label{dNL}
\\
E[dN_{Rc}(t)]&=&{\rm Tr} [W_{R1c}(t+dt)]={\cal P}_{R1c}(t) dt
\; ,
\label{dNR}
\end{eqnarray}
where  ${\cal P}_{Lc}(t)$ and ${\cal P}_{Rc}(t)$ are defined in Eqs.\
(\ref{PL1}) and (\ref{PR1}). 
Equations (\ref{condMEN})-(\ref{dNR}) 
are the main results of the paper.
One can use them to simulate the conditional (stochastic)
qubit dynamics under continuous quantum measurements by the SET.
We will present some simulation results in Sec.~\ref{sec:Simulation}

\section{connection to ``partially'' reduced density matrix}

Next, we show that 
the master equation of the ``partially'' reduced density matrix,
e.g., Eq.\ (2) of Ref.\ \onlinecite{Makhlin00},
can be obtained \cite{Goan03} by
taking a ``partial'' average on 
Eqs.\ (\ref{condMEN}) and (\ref{condMEN+1}). 
First, performing a full ensemble average over
the observed stochastic process on 
Eqs.\ (\ref{condMEN}) and (\ref{condMEN+1}) by replacing $E[dN_{Lc}(t)]$
and $E[dN_{Rc}(t)]$ by their expected values Eqs.\ (\ref{dNL})
and (\ref{dNR}), and setting 
$E[\rho_{N}^c(t)]=\rho_{N}(t)$,    
we obtain the master equation for the reduced density matrix
$\rho_{N}(t)$ and $\rho_{N+1}(t)$ as
%\begin{widetext}
\begin{eqnarray}
%\begin{equation}
&&\frac{d}{dt}\left(\begin{array}{c}\rho_N(t)\\
\rho_{N+1}(t)\end{array}\right)
+
\frac{i}{\hbar}\left(\begin{array}{c}
[{\cal H}_{\rm qb}, \rho_N(t)] \\[1pt]
[{\cal H}_{\rm qb}+2 E_{\rm int} \hat{n}, \rho_{N+1}(t)]
\end{array}\right)
\nonumber
\\ && \qquad \qquad =
\left(\begin{array}{cc}
-\check\Gamma_L & \check\Gamma_R \\
\check\Gamma_L & -\check\Gamma_R
\end{array}\right)
\left(\begin{array}{c}\rho_N(t)\\
\rho_{N+1}(t)\end{array}\right)
\; .
\label{MEreduced}
%\end{equation}
\end{eqnarray}
%\end{widetext}
Then to keep track of the
number of electrons $m=N_R$ that have tunneled into the drain, 
we need to identify the terms in Eq.\ (\ref{MEreduced}), which
come from Eqs.\ (\ref{condMEN}) and (\ref{condMEN+1}) 
and have effects corresponding 
to an electron tunneling through the right junction of the SET. 
Only one such term, originating from
$dN_{Rc}(t)\,[{\check\Gamma_R} \rho^c_{N+1}(t)/{{\cal P}_{Rc}(t)}]$ in 
Eq.\ (\ref{condMEN}), survives in Eq.\ (\ref{MEreduced}). It
is in the upper right corner of the matrix 
on the right hand side of Eq.\ (\ref{MEreduced}).
If $m$ electrons have
tunneled through the right junction of the SET
at time $t+dt$, then the accumulated number
of electrons in the drain at the earlier time $t$, 
due to the contribution of the {\em jump} term through the right junction,
should be $(m-1)$.
Hence, after writing out the number dependence $m$ or $(m-1)$
explicitly for the density matrix in Eq.\ (\ref{MEreduced}), 
we obtain \cite{partialm}  
the master equation for the ``partially'' reduced density
matrix as:
%\begin{widetext}
\begin{eqnarray}
%\begin{equation}
&&\frac{d}{dt}\left(\begin{array}{c}\rho_N(m,t)\\
\rho_{N+1}(m,t)\end{array}\right)
+
\frac{i}{\hbar}\left(\begin{array}{c}
[{\cal H}_{\rm qb}, \rho_N(m,t)] \\[3pt]
[{\cal H}_{\rm qb}+2 E_{\rm int} \hat{n}, \rho_{N+1}(m,t)]
\end{array}\right)
\nonumber
\\ &&\quad \quad  = 
\left(\begin{array}{c}
-\check\Gamma_L\rho_N(m,t)+ \check\Gamma_R\rho_{N+1}(m-1,t)\\
\check\Gamma_L\rho_N(m,t)-\check\Gamma_R\rho_{N+1}(m,t)
\end{array}\right)
%\left(\begin{array}{c}\rho_N(m,t)\\
%\rho_{N+1}(m,t)\end{array}\right)
\; .
\label{MEpartialm}
%\end{equation}
\end{eqnarray}
%\end{widetext}
%Since Eq.\ (\ref{MEpartialm}) is translation invariant in $m$ space, 
%we thus 
Making a Fourier transform 
$\rho_{N/N+1}(k,t)=\sum_m e^{-ikm}\rho_{N/N+1}(m,t)$ 
on Eq.\ (\ref{MEpartialm}), we find
that the resultant equation is exactly the same as Eq.\ (2) of 
Ref.\ \onlinecite{Makhlin00}.
If the sum over all possible values of $m$ is
taken on the ``partially'' reduced density matrix [i.e., tracing out
the detector states completely,  
$\rho_{N}(t)=\sum_{m} \rho_{N}(m,t)$], 
Eq.\ (\ref{MEpartialm}) then reduces to the master equation of the
reduced density matrix, Eq.\ (\ref{MEreduced}).
This procedure of
reducing Eqs.\ (\ref{condMEN}) and (\ref{condMEN+1}) to 
Eq.\ (\ref{MEpartialm})
and then to Eq.\ (\ref{MEreduced}), 
by successively disregarding information that distinguishes different
states of the detector, provides a connection between the approach of 
Refs.\ \onlinecite{Shnirman98} and \onlinecite{Makhlin00}
and the more detailed stochastic master equation used here.
To further demonstrate this connection, we analyze in the next section
an important ensemble quantity for an initial qubit state readout experiment,
$P(m,t)$, the probability distribution of
finding $m(=N_R)$ electrons that have 
tunneled into the drain in time $t$, considered in 
Ref.\ \onlinecite{Shnirman98}.

\section{Simulation results and discussions \label{sec:Simulation}}

Although the ``partially'' reduced density matrix 
approach \cite{Shnirman98,Makhlin00} can provide
the information about the initial qubit state, 
the system dynamics in this approach is still deterministic; 
i.e., this approach is still in an ensemble and time average sense.
If a measurement is made on the qubit system and the results are available,
the state or density matrix is conditioned on the measurement results. 
If the subsequent system evolution after the measurement is concerned,
the conditional or quantum trajectory approach should be employed.
In particular, to describe
the conditional dynamics of the qubit system in a single realization of
continuous measurements, which reflects the stochastic nature of
electrons tunneling through the SET junctions, we should use the
conditional, stochastic master equations (\ref{condMEN})-(\ref{dNR}).

\begin{figure}
\includegraphics[width=\linewidth]{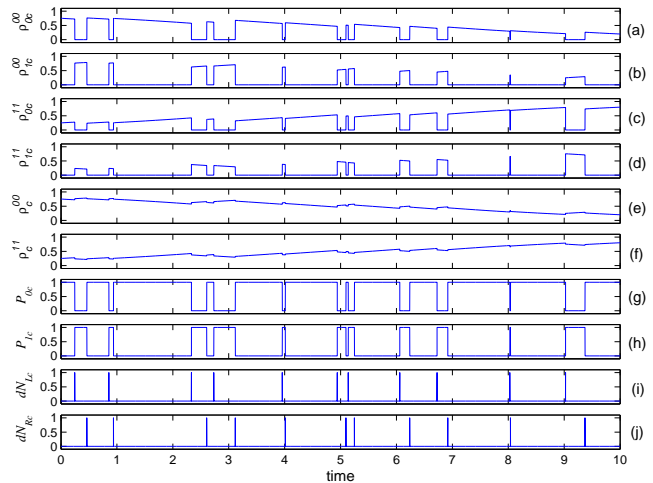}
\caption{Set of typical quantum trajectories and corresponding detection
record for the initial conditions: qubit state in
$(\sqrt{3}\,|0\rangle+|1\rangle)/2$ and the SET island in $N=0$ state. 
Other parameters are 
$E_{\rm int}=1.5 E_{\rm ch}=1500 E_J$, $\Gamma_L=(3/4)\Gamma_R=3E_{\rm
  ch}/\hbar$,
$\alpha_L=\alpha_R=0.03$ and the time is in unit of $\hbar/E_{\rm ch}$.
\label{fig:traj}}
\end{figure}

A set of typical quantum trajectories, generated using
Eqs.~(\ref{condMEN})-(\ref{dNR}), is shown in Fig.~\ref{fig:traj}(a)-(h) and
its corresponding randomly distributed moments of detections are 
presented in (i) and (j), where $\rho^{nn'}\equiv \langle n |\rho|n'\rangle$. 
Due to Coulomb blockade, 
when an electron is on the SET island,
Eqs.~(\ref{condMEN})-(\ref{dNR}) ensure that no electron can 
tunnel onto the SET island, i.e., guarantee $dN_{Lc}=0$
for the next time interval $dt$. Note that in this case, we still have
two possible measurement outcomes of either $dN_{Rc} = 1$ or $0$ in the
next time interval $dt$. But only
after a detection of $dN_{Rc}=1$ in the next or the next several time
interval(s), could another electron tunnel onto the SET island
again. This is the measurement record shown in Fig.~\ref{fig:traj}(i)
and (j). They are in the order of exactly
alternating $dN_{LC}=1$ and $dN_{Rc}=1$ time sequence. 
The conditional evolutions of the qubit alone shown in Fig.~\ref{fig:traj}(e)
and (f) can be obtained from the
sum of the joint state evolutions of (a) and (b), or 
(c) and (d),respectively. The probabilities, $P_{0/1,c}={\rm Tr}_{\rm qb}
[\rho_{0/1,c}]=\rho_{0/1,c}^{00}+\rho_{0/1,c}^{11}$,
of the SET 
island state alone in (g) and (h) can be obtained by 
summing the evolutions in (a) and (c), or (b) and (d), respectively. 
The conditional evolutions in (a)-(h) differ considerably 
from their ensemble average counterparts.

In this conditional or quantum trajectory approach, 
we are propagating in parallel the
information of the conditioned (stochastic) state evolution and 
detection record of $dN_{Lc}(t)$ and  $dN_{Rc}(t)$
in a single run of a continuous measurement process. 
One can see that in this 
approach the instantaneous system state conditions the 
measured electron tunneling events through the SET junctions
[see Eqs.\ (\ref{dNL}) and (\ref{dNR})], while the measured electron
tunneling events themselves condition 
the future evolution of the measured system [see Eqs.\
(\ref{condMEN}) and (\ref{condMEN+1})] in a self-consistent manner.
Each set of quantum trajectories (stochastic state evolutions), 
obtained from the stochastic master equations (\ref{condMEN})-(\ref{dNR}),
mimics a single history of the qubit state
in a single run of the continuous measurement experiment. 
The stochastic element in the quantum trajectories corresponds exactly
to the consequence of the random outcomes of the detection record
of the tunneling events in the SET.
The macroscopic ensemble measurement 
properties can be calculated by using large
ensembles of single electron tunneling events 
(fine grained measurement records).

If only one measurement value is recorded in
each run of experiments [for example, the number of electrons $m$ that have
tunneled into the right lead (drain) in time $t$] and
ensemble average properties 
[for example, $P(m,t)$, the probability distribution of
finding $m=(N_R)$ electrons that have tunneled through the right
junction into the drain in time $t$] are studied
\cite{Shnirman98,Makhlin00} 
over many repeated experiments, the quantum trajectory
approach will give the same result as
the master equation approach of the ``partially'' reduced 
density matrix \cite{Goan03}. 
However, more physical insight can be gained in the approach of
quantum trajectories. We demonstrate this feature below.

\begin{figure}[t]
%\centerline{\psfig{file=SET_ProbNt.eps,width=\linewidth,angle=0}}
\includegraphics[width=\linewidth]{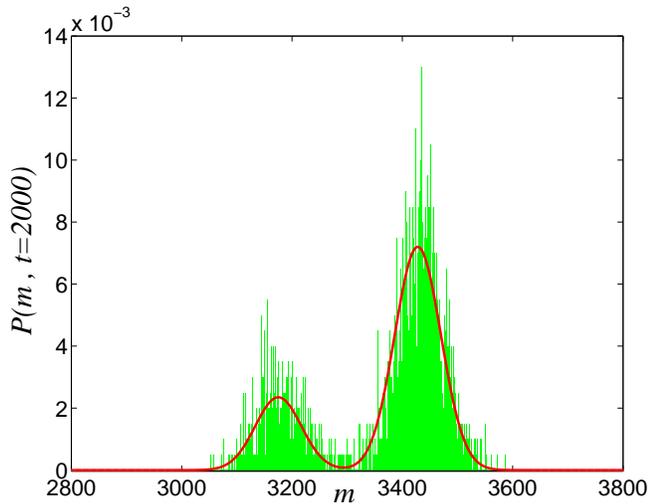}
\caption{Probability distribution $P(m,t=2000)$. The initial conditions and
  parameters are the same as those in Fig.~\ref{fig:traj}.
\label{fig:ProbNt}}
\end{figure}

We consider the case that the SCB/SET system is in the so-called Zeno
regime \cite{Shnirman98,Makhlin00,Goan01,Goan03} 
where the mixing time is much 
larger than the measurement (localization) time.  
In this regime, a good quantum readout  
measurement for an initial qubit state in the charge-state basis
can be performed by repeatedly measuring the number of electrons,
$m(t)=N_R(t)$, that 
have tunneled through the right junction into the drain of the SET 
in the same detection time period $t$. One can then use the
measurement results to construct
the probability distribution $P(m,t)$. 
%that $m(=N_R)$ electrons 
%have tunneled through the right junction into the drain during time $t$,
In Ref.~\onlinecite{Shnirman98}, the probability distribution
$P(m,t)={\rm Tr}_{\rm qb}[\rho_{N}(m,t)+\rho_{N+1}(m,t)]$ 
is obtained from the Fourier analysis of 
the partially reduced density matrix, Eq.~(\ref{MEpartialm}). The
result, obtained in this way, is plotted
in solid line in Fig.~\ref{fig:ProbNt}. 
This probability distribution $P(m,t)$ splits into two and their 
weights correspond closely to the initial qubit 
diagonal elements of $\rho^{11}(0)=0.25$ and $\rho^{00}(0)=0.75$.

In the quantum trajectory approach, $P(m,t)$
can be explicitly simulated through constructing the histogram of the
accumulated number of electrons $N_{Rc}=\sum dN_{Rc}$ up to time $t$ 
for many realizations 
of the detection records (generated together with their corresponding
quantum trajectories), and then normalizing the distribution to
one. The simulation of the normalized histogram shown 
in Fig.~\ref{fig:ProbNt} using  
$2000$ quantum trajectories and their corresponding detection records
is already in good agreement with the plot in solid line.
However, the possible individual realizations of measurement
records and their corresponding quantum trajectories [e,g., in 
Fig.~\ref{fig:traj}] do provide
insight into, and aid in the interpretation of the ensemble average
properties. This is one of the appealing features of 
the quantum trajectory approach.

For a charge qubit measured by a low-transparency 
point contact detector, this appealing feature of the quantum
trajectories is illustrated in Ref.~\onlinecite{Goan03}. 
Another advantage of the quantum trajectory approach (or Bayesian formalism) 
is that it may describe a quantum feedback
process. It has been shown \cite{Ruskov02} that one may utilize the
measurement output for the feedback control and manipulation of a qubit state.

%The difficulty in realizing the quantum trajectories experimentally
%, however, is mainly due to the requirement of large 
%bandwidth ($\geq$ GHz) of the detector circuit to record the individual
%electron tunneling signals coming out of the cryostat.
%Nevertheless, we can regard, 
%mathematically, Eqs.\ (\ref{condMEN})-(\ref{PR1}) as  
%a Monte Carlo method that allows us to model the 
%the conditional (stochastic)
%qubit dynamics under continuous quantum measurements by a SET detector.

\section{Conclusion}

To summarize, we have derived the stochastic master equation for the
SCB/SET system, which can be regarded as 
a Monte Carlo method that allows us to simulate the
continuous quantum measurement process of the SCB qubit by the SET. 
We have shown that by taking a ``partial''
average over the fine grained measurement records of the 
tunneling events in the SET, this 
stochastic master equation reduces to the master equation presented in 
Refs.\ \onlinecite{Shnirman98} and \onlinecite{Makhlin00}.
We have also presented numerical simulation for the dynamics of the
qubit in a particular realization of the readout measurement
experiment. We have shown  
that the probability distribution 
$P(m,t)$ constructed from $2000$ quantum
trajectories and their corresponding detection records, is, as
expected, in good agreement with that obtained from the Fourier analysis of 
the ``partially'' reduced density matrix.

\begin{acknowledgments}
The author would like to thank G.~J.~Milburn for his
suggestions and comments on the manuscript. 
Financial support from Hewlett-Packard is also acknowledged.
\end{acknowledgments}


\begin{references}
\bibitem{phaseshift} The last term, $\{-(\Gamma_L+\Gamma'_L)[\hat n, [\hat n,
e^{i\phi}W_c(t)e^{-i\phi}]]/2\}dt$,
in Eq.~(\ref{LjumpDM}) 
and that, $\{-(\Gamma_R+\Gamma'_R)[\hat n, [\hat n,
e^{-i\phi}W_c(t)e^{i\phi}]]/2\} dt$, in  
Eq.~(\ref{RjumpDM}) 
describe effects of a random phase shift 
  on the measured qubit state, caused by an electron when it tunnels
  onto or off the SET island. 
However, these two terms are traceless in the
  qubit basis and thus do not contribute to the average current
$I_{L/R}(t)=e\, E[dN_{Lc/Rc}(t)]/dt$, in Eqs.~(\ref{dNL}) and (\ref{dNR}).
\bibitem{partialm} Alternatively, Eq.\ (\ref{MEpartialm}) can be obtained
directly by evaluating Eq.\ (\ref{MEpartial}) in  $|m\rangle\, |N\rangle$, 
and $|m\rangle\, |N+1\rangle$ states. Similarly,
Eq.\ (\ref{MEreduced}) can be obtained directly by evaluating 
the master equation for $W(t)$ in $|N\rangle$ and $|N+1\rangle$ states.
\end{references}
\end{document}